\documentclass[pre]{revtex4}
\usepackage[dvips]{graphics}
\usepackage{latexsym}
\begin{document}
\title{Reconciliation of stability analysis with simulation in a bus route model}
\author{Scott A. Hill}
\affiliation{James Franck Institute and Department of Physics, University of Chicago, Chicago, Illinois 60637.}
\date{\today}
\begin{abstract}
In a recent paper (Physica A \textbf{296}, 320 (2001)), T. Nagatani
proposes a time-headway model for buses on a bus route, and studies
the stability of that model's homogeneous solutions.  While
investigating the phase diagram gotten by varying the rate of
passenger arrival and the initial spacing of buses, he discovers a
discrepancy between the results of simulation and those of linear
stability analysis. In this paper, we reconcile this discrepancy by
noting the existence of three types of phase diagrams, and present
simulational results which confirm the stability analysis.
\end{abstract}
\maketitle

\section{Introduction}
While there has been much interest in the study of automobile
traffic~\cite{Traffic} among scientists, there have been few
corresponding studies of
buses~\cite{Buses,Nagatani-old,Nagatani-new,Nagatani}.  The dynamics
of a bus route, while having some similarities with those of general
traffic, differ due to the added interaction of buses with passengers
at designated bus stops.  A good reason for studying the dynamics of
bus routes is that they are so often unstable.  Buses are initially
spaced at regular intervals.  However, if one bus is delayed for some
reason, it will then find a larger number of passengers waiting for it
at subsequent stops, thus delaying it further.  Meanwhile, the bus
following finds fewer passengers waiting for it, allowing it to go
faster until eventually it meets up with the delayed bus.  Clusters of
three, four, or more buses have been known to form due to this
dynamic, resulting in slower and less frequent service.

In reference~\cite{Nagatani}, Nagatani presents a time-headway model
for buses which is an extension of earlier work~\cite{Nagatani-old}.
Using linear stability analysis, he is able to determine the regions
of parameter space in which the homogeneous solution (i.e., with
buses spaced evenly apart) is unstable.  However, his analysis only
agrees qualitatively with results found by direct simulation of the
model.

In this paper, we make a slight modification to Nagatani's model, and
discover that one can reconcile the simulation results with that of
stability analysis.  On doing the analysis, one finds that there are
three different types of phase diagrams possible: one which agrees
with Nagatani's theoretical predictions, a second which agrees with
his simulation results, and a third which he does not see.  We present
our own simulational data for these three separate cases, showing
quantitative agreement.

\section{Model}
We consider the following model~\cite{Nagatani} of buses on a bus
route.  Bus stops are labelled by $s=1,2,\dots$ where stops $s$ and
$s+1$ are a distance $L$ apart.  There are $J$ buses, $j=1,\dots,J$,
which travel from stop to stop, with bus $j=J$ in the lead and bus
$j=1$ in the rear.  Every bus stops at every stop, and buses do not
pass one another.  The arrival time $t_{j,s}$ of bus $j$ at stop $s$
is given by
\begin{equation}\label{arrival-times}
t_{j,s}=t_{j,s-1}+{L\over V_{j,s-1}}+\lambda\gamma\Delta t_{j,s-1},
\end{equation}
where $\Delta t_{j,s-1}=t_{j,s-1}-t_{j+1,s-1}$ is the
\emph{time-headway}, the time gap in front of bus $j$ at stop $s-1$.
The penultimate term in Eq.~\ref{arrival-times} is the time it takes
for the bus to travel from stop to stop; $V_{j,s-1}$ is the average
velocity of the bus as it moves.  The last term is the time it takes
for passengers to board the bus at stop $s-1$: $\lambda$ is the rate of
passengers arriving at the bus stop, so $\lambda\Delta t_{j,s-1}$ is the
number of passengers that have arrived since bus $j+1$ left the stop.
The parameter $\gamma$ is the time it takes one passenger to board the
bus, and so $\lambda\gamma\Delta t_{j,s-1}$ is the amount of time needed to
board all the passengers.  One could also account for the time it
takes riding passengers to leave the bus; Nagatani shows, however,
that this is not an important effect, so we will ignore it in our own
analysis.

It is reasonable to assume that a bus driver will try to prevent
bunching by slowing down when the gap between his bus and the next is
too small.  One can model\cite{Nagatani} this discretion by writing
the average speed $V_{j,s}$ as a function $V(\Delta t_{j,s})$ of the
headway, where
\begin{equation}\label{V}
V(\Delta t)=v_{\min}+(v_{\max}-v_{\min}){\tanh[\Delta t-t_c]+\tanh t_c\over 1+\tanh t_c}.
\end{equation}
The hyperbolic tangent factor acts as a spread-out step function
centered at $t_c$, which is roughly the minimum time gap that a driver
prefers to have in front of her.  We will set $t_c=2$ in all that
follows.  The parameter $v_{\max}$ is the speed the bus would travel
with an infinite time headway.  The parameter $v_{\min}$ is currently
absent from Nagatani's formulation\cite{Nagatani}, although he did use
it in an earlier paper~\cite{Nagatani-old}.  We reintroduce it here
for generality.

It is convenient to work, not with the arrival times $t_{j,s}$, but
with the time headways $\Delta t_{j,s}$.  We thus rewrite
Eq.~\ref{arrival-times} in terms of the headways:
\begin{equation}\label{headway-eq}
\Delta t_{j,s}=\Delta t_{j,s-1}+L\left[{1\over V(\Delta t_{j,s-1})}-{1\over
V(\Delta t_{j+1,s-1})}\right]+\lambda\gamma[\Delta t_{j,s-1}-\Delta t_{j+1,s-1}].
\end{equation}

\section{Stability Analysis}

\begin{figure}[ht]
\begin{center}
\includegraphics{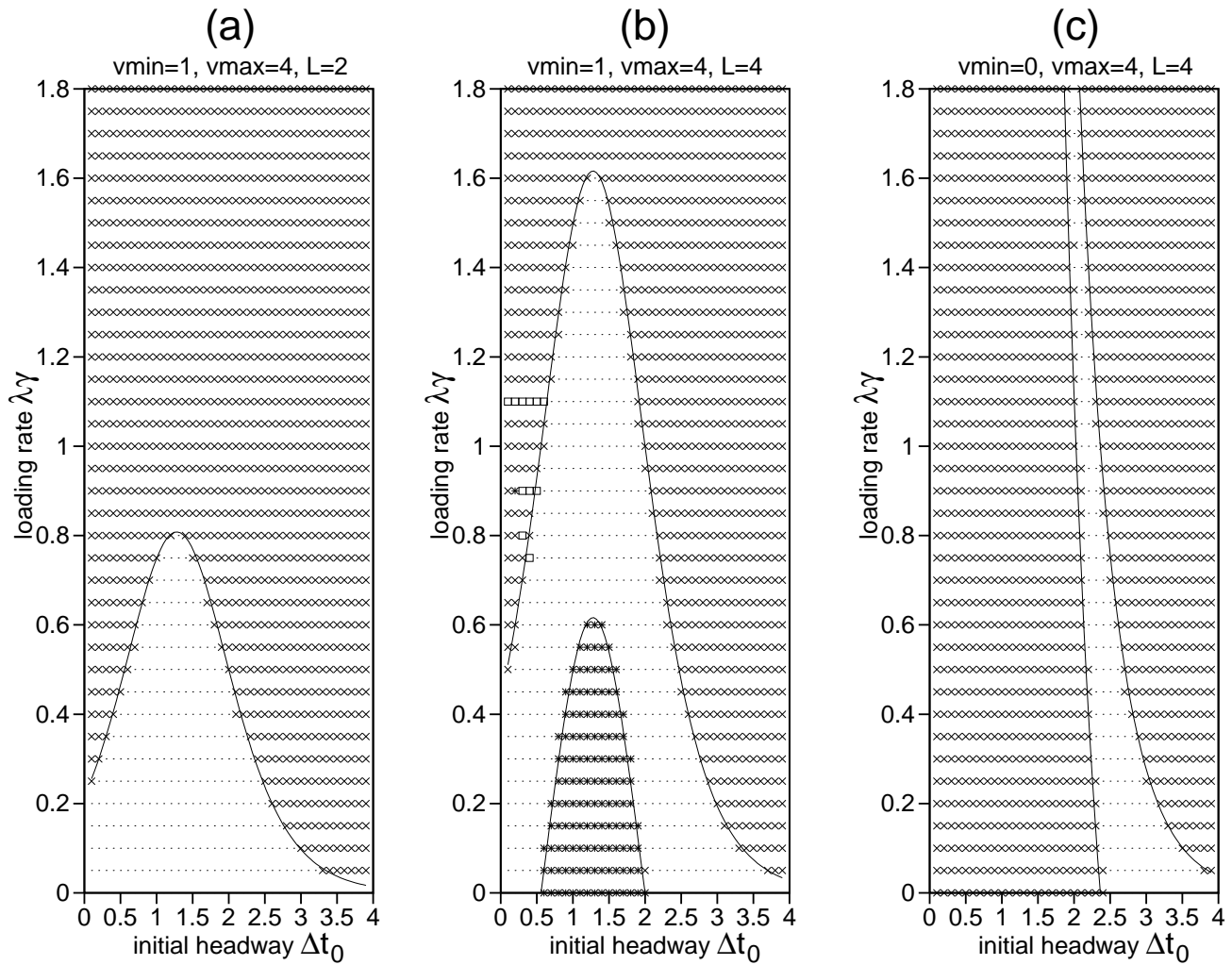}
\end{center}
\caption{\label{fig-phase} Phase diagrams for three different sets of
parameter values.  The curves are given by $L{V'(\Delta t_0)\over
V(\Delta t_0)^2}$ and $L{V'(\Delta t_0)\over V(\Delta t_0)^2}-1$,
which theory predicts will bound the stable regions of phase space.
The symbols represent simulation results: the small dots ($\cdot$)
label those simulation runs which are stable.  The crosses ($\times$)
are those runs which end up with one or more headways equal to zero.
The stars ($*$) mark those runs which end in a ``kink-jam'' phase. The
boxes ($\Box$), which only appear on the left-hand side of Figure b,
mark those runs whose headways are roughly homogeneous but higher than
the initial headway $\Delta t_0$.  (See Fig.~\ref{fig-example} for
examples of these four possible outcomes.)}
\end{figure}

We consider the stability of a homogeneous flow of buses; that is, a
situation where all buses have the same headway $\Delta t_0$.  Nagatani\cite{Nagatani}
expands Eq.~\ref{headway-eq} in terms of small deviations from
homogeneous flow, and finds that the stability condition for small
disturbances at long wavelengths is
\begin{equation}
\lambda\gamma{V(\Delta t_0)^2\over L}<V'(\Delta t_0)<(\lambda\gamma+1){V(\Delta t_0)^2\over L},
\end{equation}
where $\Delta t_0$ is the initial, constant spacing between buses, $V(\Delta
t)$ is the velocity function in Eq.~\ref{V} and $V'(\Delta t_0)$ is the
derivative of that function.  We can rewrite this in terms of the
passenger arrival rate $\lambda\gamma$,
\begin{equation}\label{inequality}
L{V'(\Delta t_0)\over V(\Delta t_0)^2}-1<\lambda\gamma<L{V'(\Delta t_0)\over V(\Delta t_0)^2},
\end{equation}
where, specifically,
\begin{equation}
{LV'(\Delta t_0)\over V(\Delta t_0)^2}
={L(v_{\max}-v_{\min})(1-\tanh t_c)(1-\tanh^2\Delta t_0)\over [v_{\min}(1-\tanh\Delta t_0)+v_{\max}(1-\tanh t_c)\tanh\Delta t_0]^2}.
\end{equation}
This inequality will allow us to construct a phase diagram
(Fig.~\ref{fig-phase}) similar to that of Fig.~8 in
Ref.~\cite{Nagatani}, where we vary the loading rate $\lambda\gamma$
and the initial spacing between buses $\Delta t_0$ .

Depending on the choice of parameters, there are actually three
general forms this phase diagram can take.  The first, corresponding
to Nagatani's analytical results, occurs when the lower bound on
$\lambda\gamma$ lies entirely below the $\Delta t_0$--axis, making it
irrelevant since the loading rate $\lambda\gamma$ is always
non-negative.  After a straightforward calculation, one can show that
the lower boundary curve reaches its maximum for that value $\Delta t_0$
which satisfies
\begin{equation}\label{topmost}
v_{\min}(1-\tanh\Delta t_0)=v_{\max}(1-\tanh t_c).
\end{equation}
From this we find that the lower boundary never rises above the
axis when
\begin{equation}\label{TypeICondition}
v_{\min}^2-[v_{\min}-v_{\max}(1-\tanh t_c)]^2>L (v_{\max}-v_{\min})(1-\tanh t_c)
\end{equation}
This type of phase diagram is seen in Figure~\ref{fig-phase}a.
When Eq.~\ref{TypeICondition} does not hold, then the lower
bound in Eq.~\ref{inequality} rises above the $\Delta t_0$ axis, and
one gets a phase diagram more like that in Figure~\ref{fig-phase}b.

Notice from Eq.~\ref{topmost} that the boundary curve has no maximum
when $v_{\min}=0$.  Figure~\ref{fig-phase}c shows the phase diagram
for this case.  The similarity between Fig.~\ref{fig-phase}c and the
simulation results in Fig.~3 of Ref.~\cite{Nagatani} is not
unexpected, since Nagatani does not use a minimum velocity in this
paper.  The puzzle is why the phase diagram of his stability analysis
is more similar to Figure~\ref{fig-phase}a; perhaps he used
$v_{\min}\ne0$ in his analysis.

\section{Simulation}
\begin{figure}[ht]
\begin{center}
\includegraphics{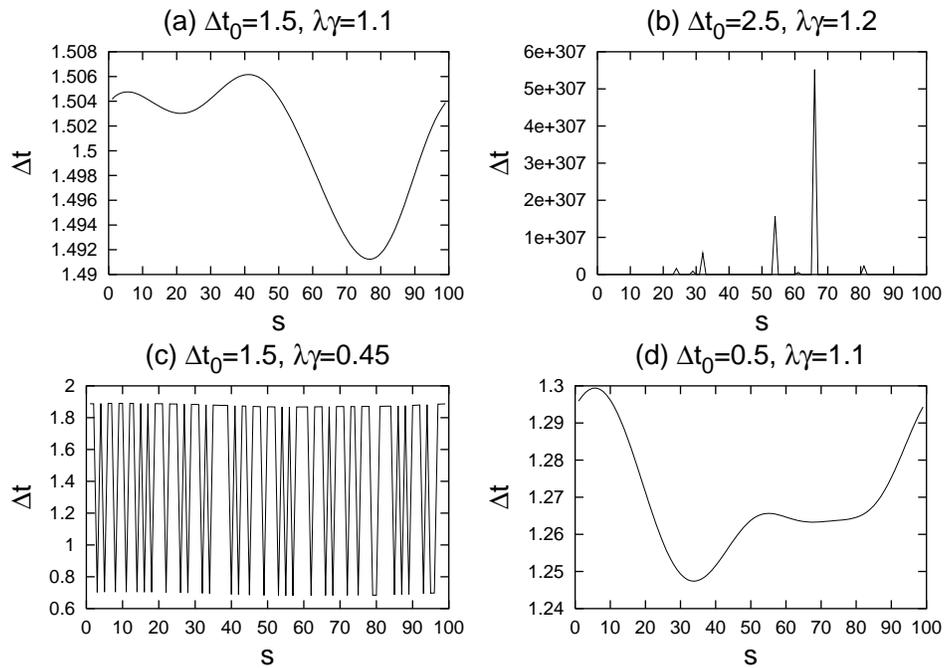}
\end{center}
\caption{\label{fig-example} Examples of possible outcomes of our
simulation.  The plot (a) shows a typical stable result, with some
fluctuation about the initial value 1.5 (in this case).  Plot (b)
shows an example of those runs with many buses bunched together; note
that this particular (not uncommon) result is completely unphysical,
and arises because we have gone far from equilibrium.  Plot (c) shows
a typical ``kink-jam'' phase; such phases are marked by having at
least one gradient $|\Delta t_{j,s}-\Delta t_{j,s-1}|$ greater than
half the distance between maximum and minimum.  The last plot, plot
(d), looks similar to the stable case, but the headways have all
drifted to be much larger than the initial value of 0.5.  To be
specific, we classified such runs as ``unstable'' if no headway came
within 0.1 of the initial headway $\Delta t_0$.}
\end{figure}

To test our theory, we evaluate Eq.~\ref{headway-eq} iteratively in
$s$.  Our initial conditions are
\begin{equation}
\Delta t_{j,0}=\Delta t_0\pm 0.1r_j
\end{equation}
where $r_j$ are random numbers chosen between $-1$ and $1$.  We use
periodic boundary conditions in bus number; so for example $\Delta
t_{N,s}=t_{N,s}-t_{1,s}$.  By stop $s=1000$, our simulations each have
one of four different outcomes: each outcome is represented by a
different symbol in Fig.~\ref{fig-phase}, and an example of each can
be found in Fig.~\ref{fig-example}.  Many choices of parameters result
in two or more buses being clumped together ($\Delta t_i=0$ for at
least one value of $i$); these are labelled by crosses in the figure.
Other runs result in what is known as a ``kink-jam''
phase\cite{Nagatani-old}, where some of the headways are much larger
than their initial values, while others are smaller.  In a few cases,
all of the headways are roughly homogeneous, but with an average value
larger than the initial value; this we also consider unstable.  When
all of the headways remain roughly where they began, we consider the
run to be stable.  We see in Fig.~\ref{fig-phase} very good agreement
between theory and simulation in all three sets of parameters.

In summary, when one takes into consideration the minimum velocity
$v_{\min}$ and the fact that there are three possible phase diagrams
for this bus system, one is able to reconcile Nagatani's simulations
with his stability analysis to a high degree.

The author would like to thank Dr. Raghuveer Parthasarathy for
indirectly encouraging this work.  This work was supported by the
Materials Research Science and Engineering Center through Grant
No. NSF DMR 9808595.
\vspace{0.5in plus 0.5in}

\end{document}